\documentclass[preprint,11pt,authoryear]{elsarticle}

\usepackage{amssymb}
\usepackage{url}
\usepackage{graphicx}
\usepackage{amsmath}
\usepackage{epstopdf}

\journal{International Journal of Forecasting}

\begin{document}

\begin{frontmatter}



\title{A Time-varying Parameter Based Seasonally-adjusted Bayesian State-space Model for Forecasting}


\author{Arnab Hazra}

\address{Indian Statistical Institute, 203, Barrackpore Trunk Road, Kolkata, India 700108.}
\address{Department of Statistics, North Carolina State University, Raleigh, United States.}
\begin{abstract}
In this paper, we develop a time-varying parameter based seasonally-adjusted Bayesian state-space model for non-stationary time series datasets where both the trend and seasonal components are present and it is the general scenario for most of the real datasets in various scientific disciplines. In spite of removing such terms using some do-and-check procedure to make the data stationary, our model directly fits a dataset and forecasts a number of future observations. For a specific prior construction we have considered, every parameter update is one-dimensional so that we don't need to invert any matrix and also we overcome the difficulty of Metropolis-Hastings steps simply by Gibbs sampling which is another advantage of this model. It can handle missing data as well which occurs very often in time series contexts. We implement it on the sufficiently large (24 years of monthly average temperature series, i.e. the number of observations $=288$) for 57 meteorological stations across India and show that for most of the cases, our method forecasts quite accurately for the months of the $25$-th year. 

\end{abstract}


\end{frontmatter}


\section{Introduction}
\label{intro}

There are plenty of scientific disciplines, e.g., the atmospheric sciences, economics, engineering, ecology, medicine etc. dealing with time-series data very frequently which are seasonally-dependent and evolving from some dynamic systems, e.g. the monthly average temperature datasets, the Consumer Price Index series etc. These kind of datasets are highly non-stationary because of the presence of trend as well as the seasonal components and the first step to deal with such datasets is to remove those components to make them stationary which is just a do-and-check procedure and also the forecasting technique is based on fitting polynomial curves to approximate such deterministic future components. A time-varying observational and evolutionary equations-based seasonally-adjusted state-space model is a possible solution to this problem. Generally, state-space models are of the following form: for $t = 1,2, \ldots$,
\begin{eqnarray}
\nonumber Y_t = f_t(X_t) + \epsilon_t \\
\nonumber X_t = g_t(X_{t-1}) + \eta_t
\end{eqnarray}
where $f_t$ and $g_t$ are generally assumed to linear; sometimes dependent on time $t$ and $\epsilon_t, \eta_t$ are assumed to be zero mean independent and identically distributed random variables; see \cite{Durbin2001} and \cite{Shumway2011} for details. The above two equations are called observational and evolutionary equations respectively. Multivariate time series are the quite obvious generalization of the univariate time series methods; see \cite{West1997} and \cite{Carvalho2007}. In a non-Bayesian framework, Kalman filter is a well-known solution to this problem. An usual state-space model do not consider the seasonal-adjustment and the time-dependency of the observational and evolutionary equations.

Jain (2001) has come up with a seasonally-adjusted state-space model but does not consider the time-varying parameter structure and hence loses the non-stationarity. Recently, \cite{Ghosh2014} has come up with a very general non-parametric and non-stationary observational and evolutionary equation based Bayesian solution but the difficulty is in inverting a $T \times T$ dimensional matrix in every Markov Chain Monte Carlo (MCMC) steps where $T$ is the length of the data which leads the method to a practical impossibility for large $T$. Actually in this approach, the latent variable does not evolve as a Markov model but depends on all it's previous stages due to the non-parametric approach and hence, a seasonal adjustment is not necessary. But, in reality, the latent variable possibly depends only on ordinary and seasonal lags and considering this structure, the computation would be actually possible in reality even for large datasets. In the illustration, the paper \cite{Ghosh2014} considers a non-seasonal dataset of size $T=25$ and the computation time is 17 hours (in C++) to generate 60,000 posterior samples as mentioned in the paper. We don't consider this model to compare our method as it is virtually impossible to implement in our set-up.

In this paper, we develop a time-varying parameter based seasonally-adjusted parametric Bayesian state-space model. In terms of computation, every parameter update is one-dimensional so that we don't need to invert any matrix and also we overcome the difficulty of Metropolis-Hastings steps and the look-up table approach of \cite{bhattacharya2007} implemented in \cite{Ghosh2014} simply by Gibbs sampling because of the closed form of the posterior distributions which is another advantage of this model. We implement it on the sufficiently large (24 years of monthly average temperature series, i.e. $T = 288$) for more than 50 stations across India and show that for most of the cases, our method forecasts quite accurately for the months of the $25$-th year. For the dataset from one station, the computation time is approximately 1 hour and 40 minutes for generating 50,000 posterior samples in R (\url{https://www.R-project.org/}).

In Section \ref{material}, we briefly overview the idea of state-space models, discuss our model and prior specification details. The posterior calculation and the Gibbs sampling steps are discussed in Section \ref{calculation}. In Section \ref{results}, we implement our method on the monthly temperature datasets and discuss the results. In Section \ref{importance}, the importance of the seasonal adjustment has been discussed. Some possible extensions are discussed in Section \ref{Future}. Section \ref{conclusions} concludes.

\section{Material and Methods}
\label{material}

The model we have chosen is as follows.

\begin{eqnarray}
Y_t &=& \beta_{0,t} + \beta_{1,t} X_t + \beta_{s,t} X_{t - 12} + \epsilon_t \\
X_t &=& \tilde{\beta}_{0,t} + \tilde{\beta}_{1,t} X_{t - 1} + \tilde{\beta}_{s,t} X_{t - 12} + \eta_t \\
X_0 &\sim& N(\mu_0, \sigma_0^2)
\end{eqnarray}
where $\epsilon_t \sim IID~ N(0, \sigma^2_\epsilon)$, $\eta_t \sim IID~ N(0, \sigma^2_\eta), \mu_0 \sim N(\xi_0, \sigma_\mu^2); ~t=1,2,\ldots,T$, with $\tilde{\beta}_{0,0} = 0, \tilde{\beta}_{1,0} = 0.5, \tilde{\beta}_{s,0} = \tilde{\beta}_{s,1} = \cdots = \tilde{\beta}_{s,10} = 0, \tilde{\beta}_{s,11} = 0.5$ and $\beta_{0,0} = 0, \beta_{1,0} = 0.5, \beta_{s,0} = \beta_{s,1} = \cdots = \beta_{s,10} = 0, \beta_{s,11} = 0.5$. Onwards in time, the parameters are chosen to be random with the mentioned prior specifications. For the better mixing properties of the MCMC chain, instead of considering $\sigma_0^2, \sigma^2_\eta, \sigma^2_\epsilon$ to be different independent parameters, we consider them to be just different scalings of a single parameter $\tau$ such that $\sigma_0^2 = c_0 \tau^{-1}, \sigma^2_\eta = c_x \tau^{-1}, \sigma^2_\epsilon = c_y \tau^{-1}$. Similarly we choose the variance parameters of the prior distributions. The joint prior distribution of the parameters is considered to be of the following dependent structure.
\begin{flushleft}
\begin{small}
\begin{eqnarray}
\nonumber && \left[\tau, \left \lbrace \tilde{\beta}_{0,t} \right \rbrace_{t=1}^{t=T}, \left \lbrace \tilde{\beta}_{1,t} \right \rbrace_{t=1}^{t=T}, \left \lbrace \tilde{\beta}_{s,t} \right \rbrace_{t=12}^{t=T}, \left \lbrace \beta_{0,t} \right \rbrace_{t=1}^{t=T}, \left \lbrace \beta_{1,t} \right \rbrace_{t=1}^{t=T}, \left \lbrace \beta_{s,t} \right \rbrace_{t=12}^{t=T} \right] \\
\nonumber &=& \left[\tau\right] \times \left[ \left \lbrace \tilde{\beta}_{0,t} \right \rbrace_{t=1}^{t=T} | \tau \right] \times \left[ \left \lbrace \tilde{\beta}_{1,t} \right \rbrace_{t=1}^{t=T} | \tau \right] \times \left[ \left \lbrace \tilde{\beta}_{s,t} \right \rbrace_{t=12}^{t=T} |\tau \right] \\
\nonumber && \times \left[ \left \lbrace \beta_{0,t} \right\rbrace_{t=1}^{t=T} | \tau \right] \times \left[ \left \lbrace \beta_{1,t} \right \rbrace_{t=1}^{t=T} | \tau \right] \times \left[ \left \lbrace \beta_{s,t} \right \rbrace_{t=12}^{t=T} | \tau \right] \\
\nonumber &=&  \left[\tau\right] \times \left \lbrace \prod_{t=1}^{T} \left[\tilde{\beta}_{0, t} | \tilde{\beta}_{0, t-1}, \tau \right] \right \rbrace \times \left \lbrace \prod_{t=1}^{T} \left[\tilde{\beta}_{1, t} | \tilde{\beta}_{1, t-1}, \tau \right] \right \rbrace \times \left \lbrace \prod_{t=12}^{T} \left[\tilde{\beta}_{s, t} | \tilde{\beta}_{s, t-1}, \tau \right] \right \rbrace \\
\nonumber && \times  \left \lbrace \prod_{t=1}^{T} \left[\beta_{0, t} | \beta_{0, t-1}, \tau \right] \right \rbrace \times \left \lbrace \prod_{t=1}^{T} \left[\beta_{1, t} | \beta_{1, t-1}, \tau \right] \right \rbrace \times \left \lbrace \prod_{t=12}^{T} \left[\beta_{s, t} | \beta_{s, t-1}, \tau \right] \right \rbrace.
\end{eqnarray}
\end{small}
\end{flushleft}

The prior specifications for the parameters are as follows.
\begin{eqnarray}
\nonumber [\tau] &\sim& Gamma(a, b) \\
\nonumber [\tilde{\beta}_{0,t}| \tilde{\beta}_{0, t-1}, \tau] &\sim& N(\tilde{\beta}_{0, t-1},  c_{\tilde{\beta}_0} t^{-2} \tau^{-1}) \\
\nonumber [\tilde{\beta}_{1,t}| \tilde{\beta}_{1, t-1}, \tau] &\sim& N(\tilde{\beta}_{1, t-1},  c_{\tilde{\beta}_1} t^{-2} \tau^{-1}) \\
\nonumber [\tilde{\beta}_{s,t}| \tilde{\beta}_{s, t-1}, \tau] &\sim& N(\tilde{\beta}_{s, t-1},  c_{\tilde{\beta}_s} t^{-2} \tau^{-1}) \\
\nonumber [\beta_{0,t}| \beta_{0, t-1}, \tau] &\sim& N(\beta_{0, t-1},  c_{\beta_0} t^{-2} \tau^{-1}) \\
\nonumber [\beta_{1,t}| \beta_{1, t-1}, \tau] &\sim& N(\beta_{1, t-1},  c_{\beta_1} t^{-2} \tau^{-1}) \\
\nonumber [\beta_{s,t}| \beta_{s, t-1}, \tau] &\sim& N(\beta_{s, t-1},  c_{\beta_s} t^{-2} \tau^{-1})
\end{eqnarray}
The priors of the parameters $\tilde{\beta}_{0,t}, \tilde{\beta}_{1,t}, \tilde{\beta}_{s,t}, \beta_{0,t}, \beta_{1,t}, \beta_{s,t}$ are specified to be dependent on time in order to ensure the convergence of the marginal variances given $\tau$; e.g. $var(\beta_{0,t}) = c_{\beta_0} \tau^{-1} \sum_{s=1}^{t} \frac{1}{s^2}$ which converges as $t \rightarrow \infty$. Also, given $\beta_{0, t-1}$, the parameter $\beta_{0,t}$ is expected to be close to $\beta_{0, t-1}$. The values of the scaling factors are chosen as $a=0.01, b=0.01, c_{\mu} = 100, c_0 = 100, c_{\tilde{\beta}_0} = 100, c_{\tilde{\beta}_1} = 1, c_{\tilde{\beta}_s} = 1, c_{\beta_0} = 100, c_{\beta_1} = 1, c_{\beta_s} = 1, c_x = 200, c_y = 200$. It is possible that these parameters need to be tuned for some particular dataset but we keep them fixed for all the time series we analyze but it is important to keep the $c_{\tilde{\beta}_1}, c_{\tilde{\beta}_s}, c_{\beta_1}, c_{\beta_s}$ to be small otherwise the larger values of these parameters may lead to breakage of the MCMC chain before convergence. The reader should note that the usual parameter range of $(-1,1)$ for the slope parameters in case of non-time-varying dynamic model is not necessary in our case as some of the slope parameters can lie outside the range of $(-1,1)$ and the time series still do not explode. To set the value for $\xi_0$, it is better if we have some prior idea about the data. For example, we deal with monthly temperature datasets across India and have prior idea that the monthly average temperature across India remains more or less around $25^{0}C - 30^{0}C$ and choose the scale parameters to be 1 and the intercept parameters to be zero. Otherwise, we can fix $\xi_0$ to be $\bar{Y} = (Y_1 + \cdots + Y_T) / T$ with the the scale parameters to be 1 and the intercept parameters to be zero which leads to an empirical Bayes approach but it also leads to the automation of the implementation of our model to any dataset in case we want a limited-length MCMC chain. Otherwise, the process will remain automatic but the convergence may need the MCMC chain to be run for a large number of times. We have adopted the second approach in this paper. In case some observations are missing, we treat it as an unknown parameter and replace $Y_T$ by $Y^{\ast}_t$ for $t=1, \ldots, T$ and update in the MCMC chain.

\subsection{Forecasting technique}

Our main aim of this modeling approach lies in the forecasting some future observations and hence we skip the issues regarding the model properties, identifiability, convergence of the parameters, simulation studies etc. Here we assume that the observations are available at $T$ time points, i.e. $Y_1, Y_2, \cdots, Y_T$ and suppose we want to predict $Y_{T+1}, \cdots, Y_{T+M}$. The obvious approach to do this is to simulate samples from the conditional distribution of $Y_{T+1}, \cdots, Y_{T+M}$ given $Y_1, Y_2, \cdots, Y_T$, i.e. the posterior predictive distribution and it can be written in the following way.
\begin{small}
\begin{flushleft}
\begin{eqnarray}
\nonumber && [Y_{T+1}, \cdots, Y_{T+M} | Y_1, \cdots, Y_T] \\
\nonumber &=& \int \cdots \int \left[Y_{T+1}, \cdots, Y_{T+M} , \tau, \mu_0, X_0, \left \lbrace\tilde{\beta}_{0,t} \right \rbrace_{t=1}^{T+M}, \left \lbrace \tilde{\beta}_{1,t} \right \rbrace_{t=1}^{T+M}, \left \lbrace\tilde{\beta}_{s, t} \right \rbrace_{t=1}^{T}, \right. \\
\nonumber && \left. \left \lbrace \beta_{0,t} \right \rbrace_{t=1}^{T+M}, \left \lbrace \beta_{1, t} \right \rbrace_{t=1}^{T+M}, \left \lbrace \beta_{s,t} \right \rbrace_{t=1}^{T+M},  \left \lbrace X_t \right \rbrace_{t=1}^{T+M}   | Y_1, \cdots, Y_T \right] d\tau \cdots d\beta_{s,T+M} \\
\nonumber &=& \int \cdots \int \prod_{t=T+1}^{T+M} \left \lbrace  \left[\tilde{\beta}_{0, t} | \tilde{\beta}_{0, t-1}, \tau \right] \times  \left[\tilde{\beta}_{1, t} | \tilde{\beta}_{1, t-1}, \tau \right] \times \left[\tilde{\beta}_{s, t} | \tilde{\beta}_{s, t-1}, \tau \right] \right. \\ 
\nonumber &&  \times \left[X_{t} | X_{t-1}, X_{t-12}, \tilde{\beta}_{0, t}, \tilde{\beta}_{1, t}, \tilde{\beta}_{s, t}, \tau \right] \times  \left[\beta_{0, t} | \beta_{0, t-1}, \tau \right] \\
\nonumber && \left. \times  \left[\beta_{1, t} | \beta_{1, t-1}, \tau \right] \times  \left[\beta_{s, t} | \beta_{s, t-1}, \tau \right] \times \left[Y_{t} | X_{t}, X_{t-12}\beta_{0, t}, \beta_{1, t}, \beta_{s, t}, \tau \right] \right\rbrace \\
\nonumber && \times \left[\tau, \left \lbrace \tilde{\beta}_{0,t} \right \rbrace_{t=1}^{t=T}, \left \lbrace \tilde{\beta}_{1,t} \right \rbrace_{t=1}^{t=T}, \left \lbrace \tilde{\beta}_{s,t} \right \rbrace_{t=12}^{t=T}, \left \lbrace \beta_{0,t} \right \rbrace_{t=1}^{t=T}, \left \lbrace \beta_{1,t} \right \rbrace_{t=1}^{t=T}, \left \lbrace X_t \right \rbrace_{t=1}^{T}, \right. \\
\nonumber && \left. \left \lbrace \beta_{s,t} \right \rbrace_{t=12}^{t=T} | Y_1, \cdots, Y_T \right] d\tau d\mu_0 \cdots d\beta_{s,T+M}.
\end{eqnarray}
\end{flushleft}
\end{small}

Thus, once we have samples from the posterior distribution of the unknown parameters and latent variable, we can generate samples from $Y_{T+1}, \cdots, \newline Y_{T+M}$ in the following way. Suppose, we have samples from the posterior distribution of $\tau, \left \lbrace \tilde{\beta}_{0,t} \right \rbrace_{t=1}^{t=T}, \left \lbrace \tilde{\beta}_{1,t} \right \rbrace_{t=1}^{t=T}, \left \lbrace \tilde{\beta}_{s,t} \right \rbrace_{t=12}^{t=T}, \left \lbrace \beta_{0,t} \right \rbrace_{t=1}^{t=T}, \left \lbrace \beta_{1,t} \right \rbrace_{t=1}^{t=T}, \left \lbrace \beta_{s,t} \right \rbrace_{t=12}^{t=T}, \newline \left \lbrace X_t \right \rbrace_{t=1}^{T}$  and denote the samples as $\tau^{\ast}, \cdots, X_T^{\ast}$. Plugging them, we draw a sample $\beta_{0,T+1}^{\ast}$ from $\left[\beta_{0, T+1} | \beta_{0, T}^{\ast}, \tau^{\ast} \right]$ and similarly for the other parameters, $X_{T+1}$ and $Y_{T+1}$ and continuing this process for $M$ steps, we can generate a set of samples $Y_{T+1}^{\ast}, \cdots, Y_{T+M}^{\ast}$ from $[Y_{T+1}, \cdots, Y_{T+M} | Y_1, \cdots, Y_T]$. The forecast values are simply some measure of the central tendency of such posterior samples and the confidence intervals of certain probability levels can be built with quantiles of such samples. In this paper, we consider the sample median of such posterior samples as the forecast values and the 0.025-th and 0.975-th sample quantiles as the 95\% confidence interval of the posterior predictive distribution.

\section{Calculation}
\label{calculation}

In our approach, the latent process $X_t$ is unobservable and hence needs to be imputed in every MCMC steps and thus, treated simply as unknown parameters. The joint posterior is our case is given by
\begin{small}
\begin{eqnarray}
\nonumber &&  \left[\tau\right] \times \left[ \mu_0 | \tau\right] \times \left[X_0 | \mu_0, \tau \right] \times  \prod_{t=1}^{11} \left \lbrace \left[\tilde{\beta}_{0, t} | \tilde{\beta}_{0, t-1}, \tau \right] \times \left[\tilde{\beta}_{1, t} | \tilde{\beta}_{1, t-1}, \tau \right] \right.  \\
\nonumber && \left. \times \left[X_{t} | X_{t-1}, \tilde{\beta}_{0, t}, \tilde{\beta}_{1, t}, \tau \right] \times  \left[\beta_{0, t} | \beta_{0, t-1}, \tau \right] \times  \left[\beta_{1, t} | \beta_{1, t-1}, \tau \right] \right. \\
\nonumber && \left. \times \left[Y_{t} | X_{t}, \beta_{0, t}, \beta_{1, t}, \tau \right] \right \rbrace \times \prod_{t=12}^{T} \left \lbrace  \left[\tilde{\beta}_{0, t} | \tilde{\beta}_{0, t-1}, \tau \right] \times  \left[\tilde{\beta}_{1, t} | \tilde{\beta}_{1, t-1}, \tau \right] \right. \\ 
\nonumber && \left. \times  \left[\tilde{\beta}_{s, t} | \tilde{\beta}_{s, t-1}, \tau \right]  \times \left[X_{t} | X_{t-1}, X_{t-12}, \tilde{\beta}_{0, t}, \tilde{\beta}_{1, t}, \tilde{\beta}_{s, t}, \tau \right] \times  \left[\beta_{0, t} | \beta_{0, t-1}, \tau \right]  \right. \\
\nonumber && \left. \times  \left[\beta_{1, t} | \beta_{1, t-1}, \tau \right] \times  \left[\beta_{s, t} | \beta_{s, t-1}, \tau \right] \times \left[Y_{t} | X_{t}, X_{t-12}\beta_{0, t}, \beta_{1, t}, \beta_{s, t}, \tau \right] \right \rbrace.
\end{eqnarray}
\end{small}

To update the parameters in a MCMC chain, we need to sample from their posterior distributions and for each of the parameters, we notice that the posterior distribution, conditioned on all other parameters, are in univariate closed forms. The reader should notice that the terms $\tilde{\beta}_{0,t}, \tilde{\beta}_{1,t}, \beta_{0, t}, \beta_{1, t} = 0$ if $t> T$ and $\tilde{\beta}_{s,t}, \tilde{\beta}_{s,t} = 0$ if $t<12$ or $t>T$. Because of the closed forms, we update them one at a time using Gibbs sampling and thus, we can avoid more complex ideas, e.g. the Metropolis-Hastings algorithm and the look-up table approach of \cite{bhattacharya2007}. The posterior distributions are as follows.

\begin{scriptsize}
\noindent\rule[0.5ex]{\linewidth}{0.1pt}
Posterior distributions of the parameters and the latent process conditioned on other parameters. \\
\noindent\rule[0.5ex]{\linewidth}{0.1pt}
\begin{eqnarray}
\nonumber [\tau] &\sim& Gamma \left(a + 4T - 10, b + \frac{1}{2} \left[ c_{\mu}^{-1}  {(\mu_0 - \xi_0)}^2  + c_0^{-1} {(X_0 - \mu_0)}^2 + c_{\tilde{\beta}_0}^{-1}  \sum_{t=1}^{T} t^2 {(\tilde{\beta}_{0,t} - \tilde{\beta}_{0, t-1})}^2 \right. \right.  \\ 
\nonumber && \left.   + c_{\tilde{\beta}_1}^{-1}  \sum_{t=1}^{T} t^2 {(\tilde{\beta}_{1,t} - \tilde{\beta}_{1, t-1})}^2 + c_{\tilde{\beta}_s}^{-1}  \sum_{t=12}^{T} {(t-11)}^2 {(\tilde{\beta}_{s,t} - \tilde{\beta}_{s, t-1})}^2 + c_{\beta_0}^{-1}  \sum_{t=1}^{T} t^2 {(\beta_{0,t} - \beta_{0, t-1})}^2 \right.  \\
\nonumber && \left. + c_{\beta_1}^{-1}  \sum_{t=1}^{T} t^2 {(\beta_{1,t} - \beta_{1, t-1})}^2 +  c_{\beta_s}^{-1}  \sum_{t=12}^{T} {(t-11)}^2 {(\beta_{s,t} - \beta_{s, t-1})}^2 \right. \\
\nonumber && \left. + c_x^{-1} \sum_{t=1}^{T} {(X_t - \tilde{\beta}_{0,t} - \tilde{\beta}_{1,t} X_{t - 1} - \tilde{\beta}_{st} X_{t - 12})}^2 \left.  + c_y^{-1} \sum_{t=1}^{T} {(Y_t - \beta_{0,t} - \beta_{1,t} X_t - \beta_{s, t} X_{t - 12})}^2 \right]  \right) \\
\nonumber [\mu_0] &\sim& N({(c_{\mu}^{-1} + c_0^{-1})}^{-1} (c_{\mu}^{-1} \xi_0 + c_0^{-1} X_0), {(c_{\mu}^{-1} + c_0^{-1})}^{-1} \tau^{-1}) \\
\nonumber [X_0] &\sim& N({(c_0^{-1} + c_x^{-1}(\tilde{\beta}_{1,1}^2 + \tilde{\beta}_{s,12}^2) + c_y^{-1}\beta_{s, 12}^2)}^{-1} (c_0^{-1} \mu_0 + c_x^{-1} \tilde{\beta}_{1,1}(X_1 - \tilde{\beta}_{0,1}) \\
\nonumber && + c_x^{-1} \tilde{\beta}_{s,12} (X_{12} - \tilde{\beta}_{0, 12} - \tilde{\beta}_{1, 12} X_{11}) + c_y^{-1} \beta_{s,12} (Y_t - \beta_{0, 12} - \beta_{1, 12} X_{12})) , \\
\nonumber && {(c_0^{-1} + c_x^{-1}(\tilde{\beta}_{1,1}^2 + \tilde{\beta}_{s,12}^2) + c_y^{-1}\beta_{s, 12}^2)}^{-1} \tau^{-1} ) \\
\nonumber [X_t] &\sim& N({(c_x^{-1} (1 + \tilde{\beta}_{1, t+1}^2 + \tilde{\beta}_{s, t+12}^2) + c_y^{-1}(\beta_{1,t}^2 + \beta_{s, t+12}^2))}^{-1} (c_x^{-1} ((\tilde{\beta}_{0, t} + \tilde{\beta}_{1,t} X_{t-1} + \tilde{\beta}_{s,t} X_{t-12}) \\
\nonumber && +  \tilde{\beta}_{1, t+1} (X_{t + 1} - \tilde{\beta}_{0, t+1} - \tilde{\beta}_{s,t+1} X_{t-11}) + \tilde{\beta}_{s, t+12} (X_{t+12} - \tilde{\beta}_{0, t + 12} - \tilde{\beta}_{1, t + 12} X_{t + 11}) ) \\
\nonumber && + c_y^{-1} (\beta_{1, t} (Y_{t} - \beta_{0,t} - \beta_{s,t}X_{t-12}) + \beta_{s,t+12} (Y_{t + 12} - \beta_{0, t+12} - \beta_{1, t+12} X_{t+12}))), \\
\nonumber && {(c_x^{-1} (1 + \tilde{\beta}_{1, t+1}^2 + \tilde{\beta}_{s, t+12}^2) + c_y^{-1}(\beta_{1,t}^2 + \beta_{s, t+12}^2))}^{-1} \tau^{-1}) \\
\nonumber [\tilde{\beta}_{0,t}] &\sim& N({(c_{\tilde{\beta}_0}^{-1} (t^2 + {(t+1)}^2) + c_x^{-1})}^{-1}(c_{\tilde{\beta}_0}^{-1} (t^2 \tilde{\beta}_{0, t-1} + {(t+1)}^2 \tilde{\beta}_{0, t+1}) \\ 
\nonumber &&  + c_x^{-1} (X_t - \tilde{\beta}_{1, t} X_{t - 1} - \tilde{\beta}_{s,t} X_{t - 12})), {(c_{\tilde{\beta}_0}^{-1} (t^2 + {(t+1)}^2) + c_x^{-1})}^{-1} \tau^{-1}) \\
\nonumber [\tilde{\beta}_{1,t}] &\sim& N({(c_{\tilde{\beta}_1}^{-1} (t^2 + {(t+1)}^2) + c_x^{-1} X_{t-1}^2 )}^{-1}(c_{\tilde{\beta}_1}^{-1} (t^2 \tilde{\beta}_{1, t-1} + {(t+1)}^2 \tilde{\beta}_{1, t+1}) \\ 
\nonumber &&  + c_x^{-1} X_{t-1} (X_t - \tilde{\beta}_{0, t} - \tilde{\beta}_{s,t} X_{t - 12})), {(c_{\tilde{\beta}_1}^{-1} (t^2 + {(t+1)}^2) + c_x^{-1} X_{t-1}^2)}^{-1} \tau^{-1}) \\
\nonumber [\tilde{\beta}_{s,t}] &\sim& N({(c_{\tilde{\beta}_s}^{-1} ({(t-11)}^2 + {(t-10)}^2) + c_x^{-1} X_{t-12}^2 )}^{-1}(c_{\tilde{\beta}_s}^{-1} ({(t-11)}^2 \tilde{\beta}_{s, t-1} + {(t-10)}^2 \tilde{\beta}_{s, t+1}) \\ 
\nonumber &&  + c_x^{-1} X_{t-12} (X_t - \tilde{\beta}_{0, t} - \tilde{\beta}_{1,t} X_{t - 1})), {(c_{\tilde{\beta}_s}^{-1} ({(t-11)}^2 + {(t-10)}^2) + c_x^{-1} X_{t-12}^2)}^{-1} \tau^{-1}) \\ 
\nonumber [\beta_{0,t}] &\sim& N({(c_{\beta_0}^{-1} (t^2 + {(t+1)}^2) + c_y^{-1})}^{-1}(c_{\beta_0}^{-1} (t^2 \beta_{0, t-1} + {(t+1)}^2 \beta_{0, t+1}) \\ 
\nonumber &&  + c_y^{-1} (Y_t - \beta_{1, t} X_{t} - \beta_{s,t} X_{t - 12})), {(c_{\beta_0}^{-1} (t^2 + {(t+1)}^2) + c_y^{-1})}^{-1} \tau^{-1}) \\
\nonumber [\beta_{1,t}] &\sim& N({(c_{\beta_1}^{-1} (t^2 + {(t+1)}^2) + c_y^{-1} X_{t}^2 )}^{-1}(c_{\beta_1}^{-1} (t^2 \beta_{1, t-1} + {(t+1)}^2 \beta_{1, t+1}) \\ 
\nonumber &&  + c_y^{-1} X_{t} (Y_t - \beta_{0, t} - \beta_{s,t} X_{t - 12})), {(c_{\beta_1}^{-1} (t^2 + {(t+1)}^2) + c_y^{-1} X_{t}^2)}^{-1} \tau^{-1}) \\
\nonumber [\beta_{s,t}] &\sim& N({(c_{\beta_s}^{-1} ({(t-11)}^2 + {(t-10)}^2) + c_y^{-1} X_{t-12}^2 )}^{-1}(c_{\beta_s}^{-1} ({(t-11)}^2 \beta_{s, t-1} + {(t-10)}^2 \beta_{s, t+1})  \\ 
\nonumber &&  + c_y^{-1} X_{t-12} (Y_t - \beta_{0, t} - \beta_{1,t} X_t)), {(c_{\beta_s}^{-1} ({(t-11)}^2 + {(t-10)}^2) + c_y^{-1} X_{t-12}^2)}^{-1} \tau^{-1}) \\ 
\nonumber [Y^{\ast}_t] &\sim& N(\beta_{0, t} + \beta_{1, t} X_t + \beta_{s,t} X_{t-12}, c_y \tau^{-1})
\end{eqnarray}
\end{scriptsize}

\noindent\rule[0.5ex]{\linewidth}{0.1pt}

\section{Results}
\label{results}
We implement our model to the monthly average temperature datasets for 57 meteorological stations spread across India. The data have been collected by India Meteorological Department for the years 1976-2000 and are available at the website \url{http://www.imd.gov.in/section/nhac/mean/110_new.htm/}. We use the datasets for the years 1976-1999 for the purpose of model fitting and cross-validate the performance based on the forecast values (along with the 95\% confidence interval of the posterior predictive distribution) and the observed for the twelve months of the year 2000. 

\begin{center}
\begin{figure}
\includegraphics[width = 5in, height = 3.5in]{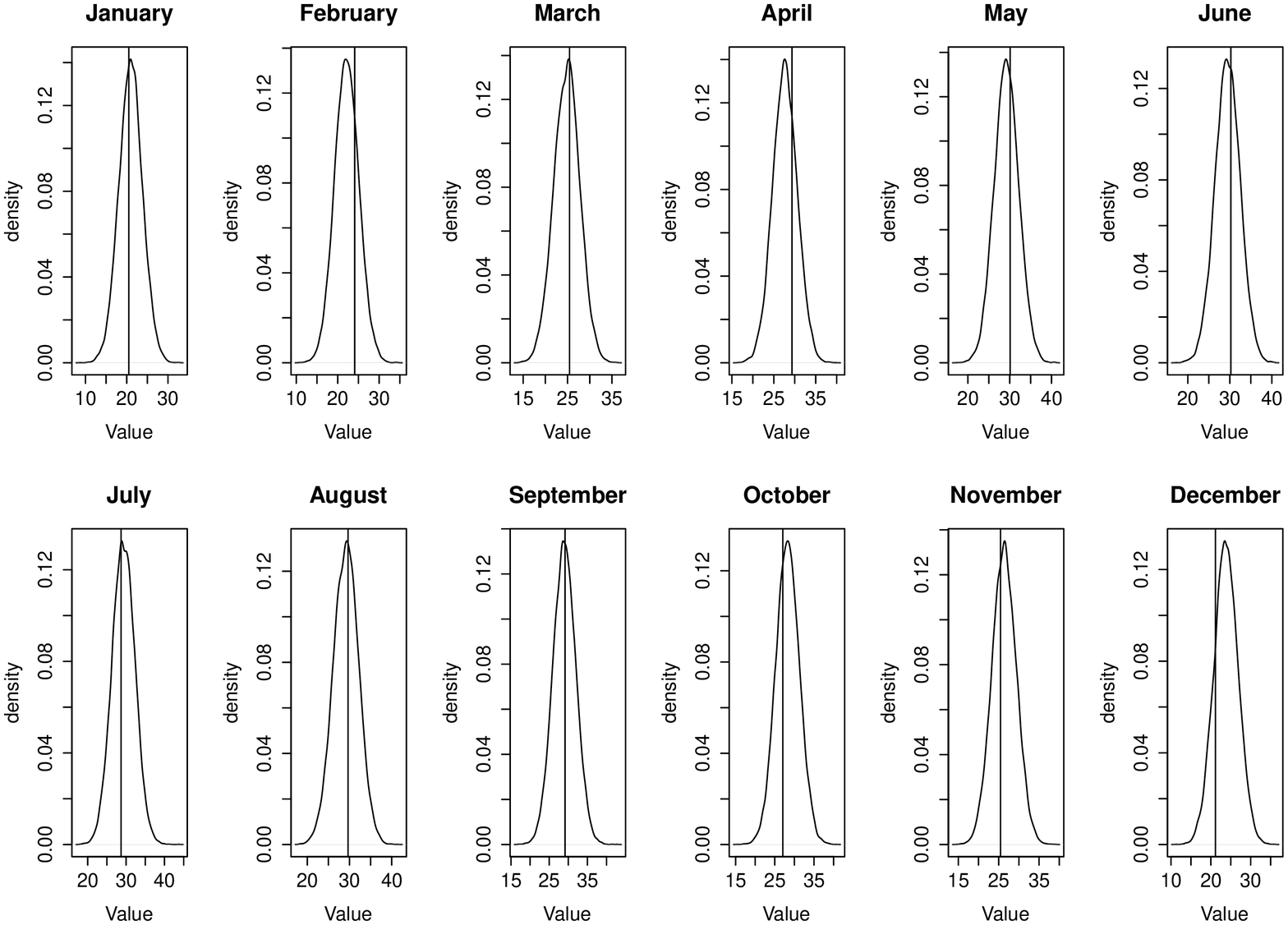}
\caption{ \small Posterior predictive distributions for 12 months of the year 2000 at Kolkata, India. The actually observed values are presented by the vertical lines.}
\label{Fig_Kolkata}
\end{figure}
\end{center}

We monitor the convergence of the MCMC chain using the trace plots. Due to huge number of parameters, we remove the first 30,000 observations as the burn-in samples and consider the next 20,000 samples to be coming from the actual posterior distribution. We do the same for all 57 stations across India. For the illustration purpose, we choose the 48-th station (Kolkata, India) and demonstrate the posterior predictive distributions for the twelve months of the year 2000 along with the actual observed value presented by the vertical line in Figure \ref{Fig_Kolkata}. We observe that the posterior central tendencies are very close to the true values for most of the months.

To study the overall performance, we check for the proportion of stations for which the actual observed values lie within the 95\% confidence intervals of the posterior predictive distribution. Except for December, 2000, the true value lies within the 95\% confidence interval for all the cases and for December, 2000, the true value lies outside the 95\% confidence interval for only one station, Jamshedpur, India.

\begin{center}
\begin{figure}
\includegraphics[width = 5in, height = 3.5in]{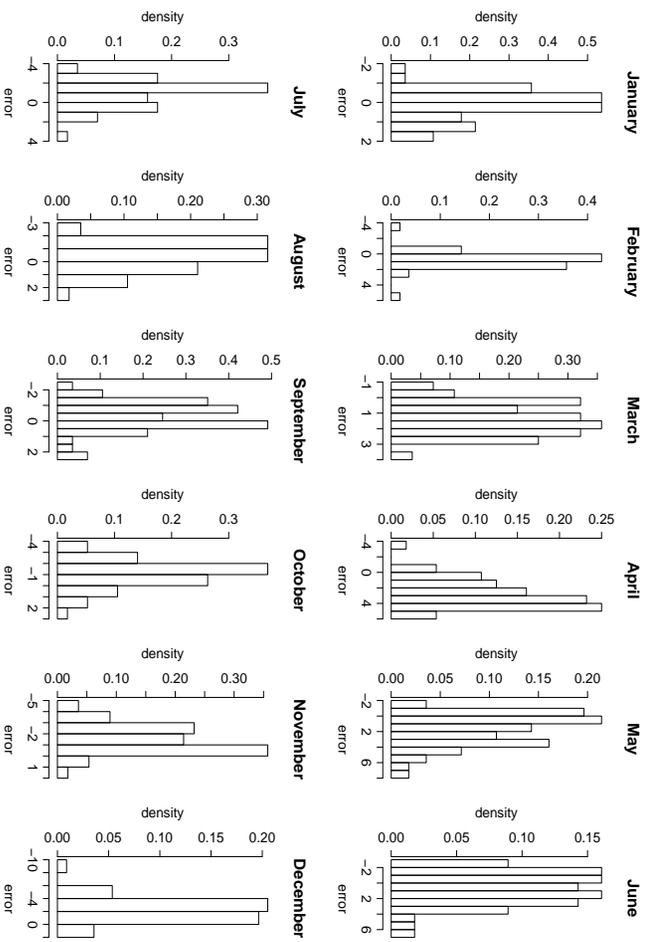}
\caption{ \small Histogram of the forecast errors for the 12 months of the year 2000.}
\label{error_hist}
\end{figure}
\end{center}

Next, for all the twelve months of the year 2000, we study the distribution of the forecasting errors, i.e. the actual observed values minus the posterior medians. The histograms are given in Figure \ref{error_hist}. We notice that the errors are closer to zero for January, February, July, August, September, October and November while we can observe high positive errors particularly for the summer months, i.e. March, April, May and June. In December, 2000, we can observe an outlier with high negative error.

\section{Importance of the seasonal adjustment}
\label{importance}

In case we do not consider the seasonal adjustment, the model would be a simple linear time-varying parameter based Bayesian state-space model, i.e. of the form

\begin{eqnarray}
Y_t &=& \beta_{0,t} + \beta_{1,t} X_t + \epsilon_t \\
X_t &=& \tilde{\beta}_{0,t} + \tilde{\beta}_{1,t} X_{t - 1} + \eta_t \\
X_0 &\sim& N(\mu_0, \sigma_0^2)
\end{eqnarray}

Here we compare the performance of our model with this existing state-space model and study the error distribution. We consider the same prior structure for the parameters $\tau, \left \lbrace \tilde{\beta}_{0,t} \right \rbrace_{t=1}^{t=T}, \left \lbrace \tilde{\beta}_{1,t} \right \rbrace_{t=1}^{t=T}, \left \lbrace \beta_{0,t} \right \rbrace_{t=1}^{t=T}, \left \lbrace \beta_{1,t} \right \rbrace_{t=1}^{t=T}$ as before. In this approach, the latent process $X_t$ is again unobservable and thus, treated simply as unknown parameters. The joint posterior is this case is given by
\begin{small}
\begin{eqnarray}
\nonumber &&  \left[\tau\right] \times \left[ \mu_0 | \tau\right] \times \left[X_0 | \mu_0, \tau \right] \times \prod_{t=1}^{T} \left \lbrace  \left[\tilde{\beta}_{0, t} | \tilde{\beta}_{0, t-1}, \tau \right] \times  \left[\tilde{\beta}_{1, t} | \tilde{\beta}_{1, t-1}, \tau \right] \right. \\ 
\nonumber && \left. \times \left[X_{t} | X_{t-1}, \tilde{\beta}_{0, t}, \tilde{\beta}_{1, t}, \tau \right] \times  \left[\beta_{0, t} | \beta_{0, t-1}, \tau \right] \times  \left[\beta_{1, t} | \beta_{1, t-1}, \tau \right] \times \times \left[Y_{t} | X_{t}, \beta_{0, t}, \beta_{1, t},  \tau \right] \right \rbrace.
\end{eqnarray}
\end{small} 
The univariate posterior distributions for all the parameters except $\tau$, remain same as in the case of our model after replacing $\tilde{\beta}_{s,t}, \beta_{s,t}$ for all $t$ by zero. The posterior distribution for the parameter $\tau$ is given by
\begin{scriptsize}
\begin{eqnarray}
\nonumber [\tau] &\sim& Gamma \left(a + 3T + 1, b + \frac{1}{2} \left[ c_{\mu}^{-1}  {(\mu_0 - \xi_0)}^2  + c_0^{-1} {(X_0 - \mu_0)}^2 + c_{\tilde{\beta}_0}^{-1}  \sum_{t=1}^{T} t^2 {(\tilde{\beta}_{0,t} - \tilde{\beta}_{0, t-1})}^2 \right. \right.  \\ 
\nonumber && \left.   + c_{\tilde{\beta}_1}^{-1}  \sum_{t=1}^{T} t^2 {(\tilde{\beta}_{1,t} - \tilde{\beta}_{1, t-1})}^2 + c_{\beta_0}^{-1}  \sum_{t=1}^{T} t^2 {(\beta_{0,t} - \beta_{0, t-1})}^2 + c_{\beta_1}^{-1}  \sum_{t=1}^{T} t^2 {(\beta_{1,t} - \beta_{1, t-1})}^2 \right.  \\
\nonumber && \left. + c_x^{-1} \sum_{t=1}^{T} {(X_t - \tilde{\beta}_{0,t} - \tilde{\beta}_{1,t} X_{t - 1})}^2 \left.  + c_y^{-1} \sum_{t=1}^{T} {(Y_t - \beta_{0,t} - \beta_{1,t} X_t)}^2 \right]  \right)
\end{eqnarray}
\end{scriptsize}

The posterior predictive distribution is given by
\begin{small}
\begin{flushleft}
\begin{eqnarray}
\nonumber && [Y_{T+1}, \cdots, Y_{T+M} | Y_1, \cdots, Y_T] \\
\nonumber &=& \int \cdots \int \prod_{t=T+1}^{T+M} \left \lbrace  \left[\tilde{\beta}_{0, t} | \tilde{\beta}_{0, t-1}, \tau \right] \times  \left[\tilde{\beta}_{1, t} | \tilde{\beta}_{1, t-1}, \tau \right] \times \left[X_{t} | X_{t-1}, \tilde{\beta}_{0, t}, \tilde{\beta}_{1, t}, \tau \right]  \right. \\ 
\nonumber && \left. \times  \left[\beta_{0, t} | \beta_{0, t-1}, \tau \right] \times  \left[\beta_{1, t} | \beta_{1, t-1}, \tau \right] \times \left[Y_{t} | X_{t}, \beta_{0, t}, \beta_{1, t}, \tau \right] \right\rbrace \times \left[\tau, \left \lbrace \tilde{\beta}_{0,t} \right \rbrace_{t=1}^{t=T}, \right.\\
\nonumber && \left. \left \lbrace \tilde{\beta}_{1,t} \right \rbrace_{t=1}^{t=T}, \left \lbrace \beta_{0,t} \right \rbrace_{t=1}^{t=T}, \left \lbrace \beta_{1,t} \right \rbrace_{t=1}^{t=T}, \left \lbrace X_t \right \rbrace_{t=1}^{T}| Y_1, \cdots, Y_T \right] d\tau d\mu_0 \cdots d\beta_{s,T+M}. \\
\nonumber && 
\end{eqnarray}
\end{flushleft}
\end{small}

Hence, once a set of samples from the posterior distribution of the unknown parameters and latent variable, we can generate samples from $Y_{T+1}, \ldots, Y_{T+M}$ in the similar way as earlier. Again, we consider the sample median of such posterior samples as the forecast values and the 0.025-th and 0.975-th sample quantiles as the 95\% confidence interval of the posterior predictive distribution.

\begin{center}
\begin{figure}
\includegraphics[width = 5in, height = 3.5in]{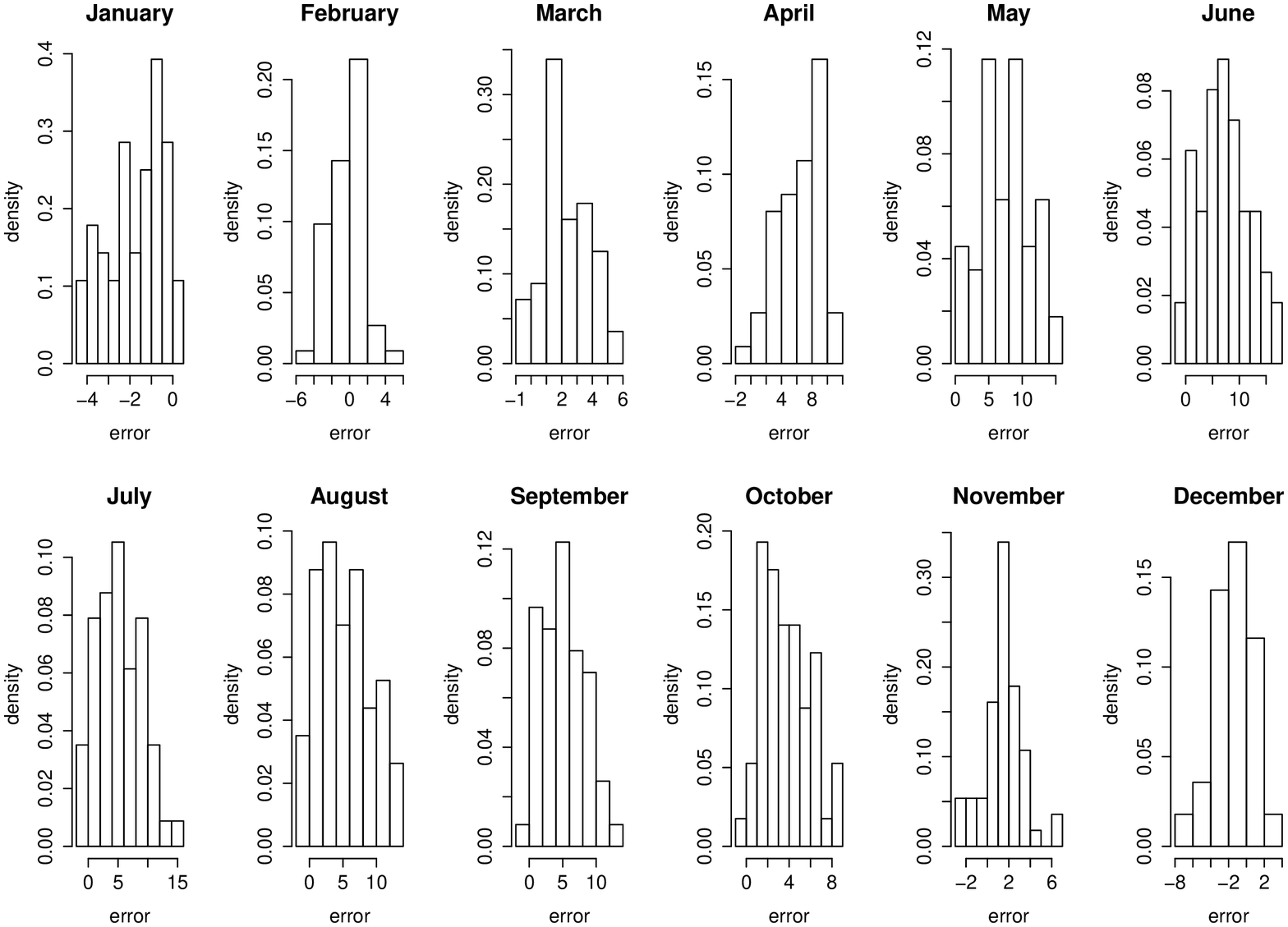}
\caption{ \small Histogram of the forecast errors for the 12 months of the year 2000 using the state-space model without seasonal-adjustment.}
\label{error_hist_HMM}
\end{figure}
\end{center}

We fit the data for the years 1976-1999 for the model fitting purpose and forecast for the twelve months of the year 2000 using this model. The histogram of the errors are provided in Figure \ref{error_hist_HMM}.

We notice that high positive errors for the months March, April, May, June, July, August, September, October and November, high negative errors for the months January and December and they are more or less distributed around zero only for the month February. Comparing Figure \ref{error_hist} and Figure \ref{error_hist_HMM}, we can notice that the absolute errors are much higher compared to the forecast of our seasonally-adjusted model.

\section{Discussions and Future Work}
\label{Future}

Compared to the forecast made by time-varying parameter based ordinary state-space model, our seasonally-adjusted model preforms better for all the months. The seasonality leads to non-linear dynamics in the monthly temperature series and the ordinary state-space model fails to capture this non-linearity. One possible solution is to consider the non-parametric approach by \cite{Ghosh2014} which is enormously time-consuming. Thus, the first approach should be to identify the nature of non-linearity present in the dataset, for example, the predominant period for a monthly temperature series is 12, for the Johnson and Johnson quarterly earnings data presented in \cite{Shumway2011}, the predominant period is 4. If the period is different from 12, our method can be simply adjusted to make it suitable for the data; for example, replacing the latent variable at a lag 12 by one at a lag equal to the predominant period. Because of the time-varying nature of the parameters, our method is non-stationary. Besides, we use the possibly easiest version of the MCMC technique, i.e. Gibbs sampling instead of more complex algorithms which often become difficult to understand and use by interdisciplinary researchers. \\

Future work of this paper is three-fold. Firstly, this work can be extended to a spatio-temporal non-stationary Bayesian modeling for seasonal spatio-temporal datasets, e.g. the monthly temperature series recorded at a number of stations throughout a country. Secondly, here we consider the latent process $X_t$ to be dependent on $X_{t-1}$ and $X_{t-12}$. It is possible that the latent process is dependent on a few more lags, e.g. $X_t$ is dependent on some ordinary lags, say $X_{t-1}, \cdots, X_{t-p}; 1 \leq p \leq 12$ and $X_{t-12}, X_{t-24}, \cdots, X_{t-12q}; q \geq 1$. Again, we can assume $p$ or $q$ to be known or unknown. In case $p$ and $q$ are known, the similar procedure as ours can be adapted with Gibbs sampling steps in a more generalized fashion while the Reversible Jump MCMC (Green, 1995) along with Gibbs sampling is a possible solution to the case when $p$ and $q$ are unknown. Thirdly, if the process is not continuous, for example, a dependent Bernoulli process, instead of considering normal observational and evolutionary equations, if we assume some generalized linear model so that the posterior distribution is no more in some closed form, the TMCMC approach by Dutta and Bhattacharya (2014) is a possible faster solution for known $p$ and $q$ while TTMCMC approach by Das and Bhattacharya (2015) is a faster solution for unknown $p$ and $q$.

\section{Conclusions}
\label{conclusions}
In case the seasonality is present in a time series dataset, the process is highly non-linear and an ordinary state-space model is not a good choice for the forecasting purpose. Besides, a real time series dataset often contains a trend component. Thus, to transform the data into a stationary process, we need to remove the trend and seasonal components. But once we are interested in forecasting, the future values of the trend components are not known and hence, they need to be approximated by curve fitting. All these steps are just do-and-check procedures and often leads to difficulties for the inter-disciplinary researchers interested in forecasting a time series for a certain future horizon.

Our method is a practical solution to this problem. The parameters of the model are time-dependent and hence, can handle non-stationarity. To fit our model to a time series, we do not need any of these steps and it is completely automatic, i.e. once we provide the time series data as an input and the number of time points where we want to forecast, our model directly returns the output with a confidence interval. Compared with the model without seasonal adjustment, our method performs much better in prediction for a seasonally-dependent time series. 

\section*{Acknowledgment}
The author would like to thank Dr. Soumendra Nath Lahiri from North Carolina State University for a number of valuable suggestions. The author would also like to thank Dr. Sourabh Bhattacharya from ISRU, Indian Statistical Institute for some discussion regarding a few important references.




\section*{Vitae}

Arnab Hazra is presently a PhD student at the Department of Statistics, North Carolina State University and his research interest is developing advanced spatio-temporal forecasting models in a Bayesian framework arising from various scientific disciplines. He has received his Bachelor of Statistics and Master of Statistics degrees from the prestigious Indian Statistical Institute where he started developing ideas about automatic model selection particularly in the meteorological context.

\end{document}